\pdfoutput=1
\documentclass[a4paper]{jpconf}
\usepackage{graphicx}
\usepackage[colorlinks = true,
            linkcolor = blue,
            urlcolor  = blue,
            citecolor = blue,
            anchorcolor = blue]{hyperref}
\usepackage{array}
\usepackage{amsmath}

\begin{document}
\title{Galactic Superbubbles in 3D: Wind Formation and Cloud Shielding}

\author{O.~Su\'arez-L\'opez$^1$, A.~S.~Villares$^1$, and W.~E.~Banda-Barrag\'an$^{1,2}$}

\address{$^{1}$ Escuela de Ciencias F\'isicas y Nanotecnolog\'ia, Universidad Yachay Tech, Hacienda San Jos\'e S/N, 100119 Urcuqu\'i, Ecuador\\
$^{2}$ Hamburger Sternwarte, University of Hamburg, Gojenbergsweg 112, 21029 Hamburg, Germany\\}

\ead{wbandabarragan@gmail.com}

\begin{abstract}
Galactic superbubbles are triggered by stellar feedback in the discs of star-forming galaxies. They are important in launching galactic winds, which play a key role in regulating the mass and energy exchange in galaxies. Observations can only reveal projected information and the 3D structure of such winds is quite complex. Therefore, numerical simulations are required to further our understanding of such structures. Here, we describe hydrodynamical simulations targeting two spatial scales. Large-scale superbubble models reveal supernova-driven outflows, and their subsequent merging, which leads to galactic wind formation. Additionally, the turbulence parameter $\sigma_t$ not only affects disc formation, but also influences mass and energy characteristics, controlling gas distribution and the injection rate in the simulated star formation zone. Small-scale wind-multicloud models indicate that isolated clouds are susceptible to instabilities, leading to fragmentation and dense gas destruction. In contrast, in closer cloud configurations, the condensation mechanism becomes important owing to hydrodynamic shielding, which helps to maintain the cold material throughout the evolution of the system. These simulations provide a comprehensive picture of galactic winds, showing how large-scale superbubble dynamics create the environment where small-scale wind-multicloud interactions shape the interstellar and circumgalactic media, ultimately regulating galaxy evolution.

\end{abstract}

\section{Introduction}
When addressing the evolution and formation of our Galaxy, it is important to mention the role of galactic winds as they regulate the mass and energy exchange in in star-forming galaxies (see e.g. \cite{2023MNRAS.518L..87N} \& \cite{2021ApJ...913...68Z}). Galactic winds transport metals from the Interstellar Medium (ISM) to the Circumgalactic Medium (CGM) and out of the Galaxy, then affecting the chemical composition and evolution of the halo gas in the Milky Way (see \cite{heckman2017galactic} \& \cite{2016ApJ...833...54W}). These winds occur in starburst-like systems, where the energy, momentum, and radiation produced by thousands of supernova (SN) explosions (e.g. \cite{1985Natur.317...44C,murray2005maximum}) give rise to these winds (see a recent review by \cite{2023ARA&A..61..131F}).\par

These winds are multiphase, i.e. they are composed of hot, warm and cold gas. The hot gas reaches temperatures higher than $10^7\,\rm K$ and can be observed via X-rays. Adiabatic and radiative winds are the most well-studied and compared to the canonical CC85 wind model (see \cite{2022HEAD...1910704L}). As a result of the SN explosions, the gas expands adiabatically first and then it starts to loss internal energy via radiative cooling. Thus, as some gas cools, the wind becomes a combination of hot and cold gas, which produces an increase in wind speed and Mach number \cite{heckman2017galactic}. Depending on the sizes and masses of the stars that produce the SNe, the temperature and speed of the winds can vary. For example, \cite{2024arXiv240105623B} shows that for stars between 10 and 20  $M_\odot$ the driven-winds can reach speeds of up to $\sim 10^3$ $\rm km\,s^{-1}$. \par

The Galactic Centre (GC) of our Galaxy offers a unique view of galactic outflows (e.g., see \cite{2023A&A...674L..15V}), given its proximity to us. If fact, the GC is a zone with high star formation (SF), as within this region a substantial amount of energy is released due to the death of thousands of stars. Each massive star approximately releases $3 \times 10^{51}\,\rm erg$  \cite{maoz2016astrophysics}. Furthermore, galaxies with high SF rate, known as starbursts, are also important for formulating new theories of the formation and evolution of winds in galaxies. The energies produced in those zones form a wind, which can drag material out of the disc \cite{2000MNRAS.314..511S}. Moreover, material that is pushed out of the disc can fill cavities called bubbles, these lobular structures can merge and grow even larger with time, resulting in the formation of superbubbles. As a consequence superbubbles are formed by supernova explosions and stellar winds (see \cite{2006ApJ...638..196M} \& \cite{2021A&A...646A..66P}).  

Understanding the formation and evolution of winds at various scales is crucial to shed light on the physical process that regulate baryon exchange in galaxies. It is then necessary to make more quantitative measurements, because the observational evidence of winds and superbubbles only provide projected information (in line emission) and only information along specific sight-lines (in line absorption). In this paper, we describe models of galactic wind phenomena at two spatial scales, namely $\sim 1\,\rm kpc$ via disc-wind models and $\sim 100\,\rm pc$ via wind-multicloud models. These models were presented in a contributed talk and a poster at the XVIII Encuentro de F\'isica by the authors. Our objectives are the following:

\begin{itemize}
    \item Characterise the initial density distribution of galactic wind atmospheres by analysing the gravitational potential in order to incorporate this into a numerical model.
    \item Evaluate the evolution and formation of superbubbles for different gas distributions by varying the turbulence strength parameter ($\sigma_t$).
    \item Analyse the changes in mass and energy injection for the different models studied to determine the influence of $\sigma_t$. 
    \item Characterise the evolution of wind-multicloud systems via hydrodynamical simulations of multicloud systems interacting with a hot supersonic wind gas.
    \item Evaluate the efficiency of hydrodynamic shielding, contrasting the thermodynamical and dynamical evolution of multicloud systems with varying cloud separation distances, $\delta$.
\end{itemize}

The paper is organised as follows: in Section \ref{sims} we present our numerical methods and models, in Section \ref{results} we show the main findings of our studies at the two spatial scales mentioned above, and in Section \ref{conclusions} we summarise our conclusions.



\section{Simulations}
\label{sims}

\subsection{The PLUTO software}
The models were performed in PLUTO, an open-source, publicly-available  code, reported by \cite{mignone2007pluto}. This software is useful for studying astrophysical gases as it includes numerical solvers to solve parabolic and hyperbolic partial differential equations. PLUTO requires Python and a C compiler to run the simulations as it is based on the C language. In addition, parallelisation is implemented within the code using Message Passing Interface (MPI). The PLUTO code is optimal for our simulations because it is portable and scalable to 1000's of CPUs \cite{mignone2007pluto}.

\subsection{Python}
Python\footnote{\url{https://www.python.org/}} is a versatile programming language used for scientific visualisation and data analysis thanks to its high level syntax. In addition, the scripts that python allows us to perform are indispensable for post-processing simulation data obtained via High-Performance Computing (HPC). We use python scripts to perform data analysis and image processing of simulation outputs. Python offers flexibility as it can perform on single or multi-core facilities, making data processing more effective. The plots and movies reported here were all produced with python.

\subsection{Initial and Boundary Conditions}

In the case of the formation of superbubbles it is necessary first to implement the gravitational potential, for this purpose, two contributions are taken into account according to \cite{2000MNRAS.314..511S} and \cite{2008ApJ...674..157C}, that of the spheroid (equation \ref{spheroid_pot}) and that of the disc (equation \ref{disk_pot}), the sum of the two gives us as a result the total potential ($\Phi_{tot} = \Phi_{ss} + \Phi_{disc}$), where the radius of the spheroid is $R=\left(r^2+z^2 \right)^{1/2}$,$r$ is the radial coordinate of a point,  the mass of the disc and the spheroid are $M_{disc}$ and $M_{ss}$, respectively. The radial scale length is $a$ and the vertical scale length is $b$, and $r_0$ is the core radius.

\begin{equation}
    \Phi_{ss}(R) = -\frac{G M_{ss}}{r_0}\left\{ \frac{ln \left[ (R/r_0)+ \sqrt{1+(R/r_0)^2}\right]}{(R/r_0)}\right\}
    \label{spheroid_pot}
\end{equation}
\begin{equation}
    \Phi_{disc}(r,z) = - \frac{G M_{disc}}{\sqrt{r^2 + \left (a+\sqrt{z^2 + b^2} \right)^2}}
    \label{disk_pot}
\end{equation}

Once the gravitational potential is described, the initial distribution of the gas is implemented. The ISM consists of two components: the isothermal halo and the turbulent disc. The density distribution of these components is described in equations \ref{halo_density} and \ref{disk_density}. Where $e_h$ is the ratio of the azimuthal and Keplerian velocity, and also the isothermal sound speed is ${c_{s,h}^2} = (k T_h / \mu m)^{1/2}$, for the case of the halo density distribution. On the other hand, for the distribution of density in the disc we have ${c_{s,d}^2} = (k T_d / \mu m)^{1/2}$, which is the warm gas sound speed, the ratio of same velocities as in the case of the halo $e_d$, and finally the turbulence parameter $\sigma_t$ (see \cite{2000MNRAS.314..511S} \& \cite{2008ApJ...674..157C}).

\begin{equation}
    \frac{\rho_{halo}(r,z)}{\rho_{halo}(0,0)} = exp \left[- \frac{\Phi_{tot}(r,z) - e_h^2\Phi_{tot}(r,0)-(1-e_h^2)\Phi_{tot}(0,0)}{c_{s,h}^2} \right]
    \label{halo_density}
\end{equation}

\begin{equation}
    \frac{\rho_{disc}(r,z)}{\rho_{disc}(0,0)} = exp \left[- \frac{\Phi_{tot}(r,z) - e_d^2\Phi_{tot}(r,0)-(1-e_d^2)\Phi_{tot}(0,0)}{\sigma_t^2 + c_{s,d}^2} \right]
    \label{disk_density}
\end{equation}

Thus, with all the equations described above the atmosphere was established for two different models. In the case of this project, since we consider a non-rotating halo  the value of $e_h$ is zero, and for a gaseous disc $e_d = 0.95$, these values hold for the two models to be studied, for which the initial parameters are in Table \ref{table1}. However, this gas distribution under the presence of the gravitational potential collapses towards the centre in extremely short times. To avoid this, the rotation velocity in the plane of the galaxy has been implemented by equation \ref{rotation} \cite{2000MNRAS.314..511S}.

\begin{equation}
    v_{\Phi} = \left( r \frac{\partial \Phi_{tot}}{\partial r}\right)^{1/2}
    \label{rotation}
\end{equation}

\begin{table}[htbp]
\centering
\caption{Parameters for the two superbubble formation models based on data from our Galaxy.} 
\vspace{10pt}

\begin{tabular}{>{\centering\arraybackslash}m{4cm} >{\centering\arraybackslash}m{2cm} >{\centering\arraybackslash}m{4cm}}

\hline
Parameter                  & Symbol & Value \\ \hline
Disc mass                  & $M_{disc}$   &   $1 \times 10^{10} $ ($M_{\odot}$)   \\
Spheroid mass              & $M_{ss}$  & $9 \times 10^9$ ($M_{\odot}$)     \\
Core radius                & $r_{0}$    &  250 (pc)     \\
Radial scale length        &    a    &      150 (pc) \\
Vertical scale length      &    b    &    50 (pc)   \\
Central halo density       &   $n_h$     &     0.1 ($\rm cm^{-3}$)  \\
Average disc density       &  $n_{d,avg}$   & 100 ($\rm cm^{-3}$)       \\
Halo temperature           &    $T_h$    & $5 \times 10^6$ (K)       \\
Average disc temperature   &    $T_{d,avg}$    &   $ 10^4$ (K)    \\
Star formation zone radius &    $r_{sf}$    &   150 (pc)   \\
Star formation zone height &     $h_{sf}$   &    40 (pc)  \\
Mass injection rate        &     $\dot{M}$   &    0.1 ($M_{\odot} yr^{-1}$)   \\
Energy injection rate      &    $\dot{E}$    &   $ 10^{42}$ (erg $s^{-1}$)   \\
\hline

\end{tabular}

\label{table1}
\end{table}

As a consequence of the rotation velocity, some part of the gas escapes through the edges of the computational domain and also tends to re-enter. With the aim of preventing the re-injection of gas from the edges, diode boundary conditions were implemented. In these conditions, and considering that our computational domain is a cube of aspect ratio (1:1:1) based on Cartesian coordinates ($x, y, z$), the velocity of the gas that escapes through the beginning ghost zones ($v_{X_n\_Beg}$) is changed by the minimum value of gas velocity. Moreover the gas velocity at the end ghost zones ($v_{X_n\_End}$) is changed by the maximum value of the velocity at this region (see equation \ref{boundary}).

\begin{equation}
     \begin{cases}
         v_{X_n\_Beg} & \to Min[v_{X_n}(Beg)] \\
         v_{X_n\_End} & \to Max[v_{X_n}(Max)] 
     \end{cases}
     \label{boundary}
 \end{equation}

\begin{figure}[htbp]
  \centering
  \includegraphics[width=1\linewidth]{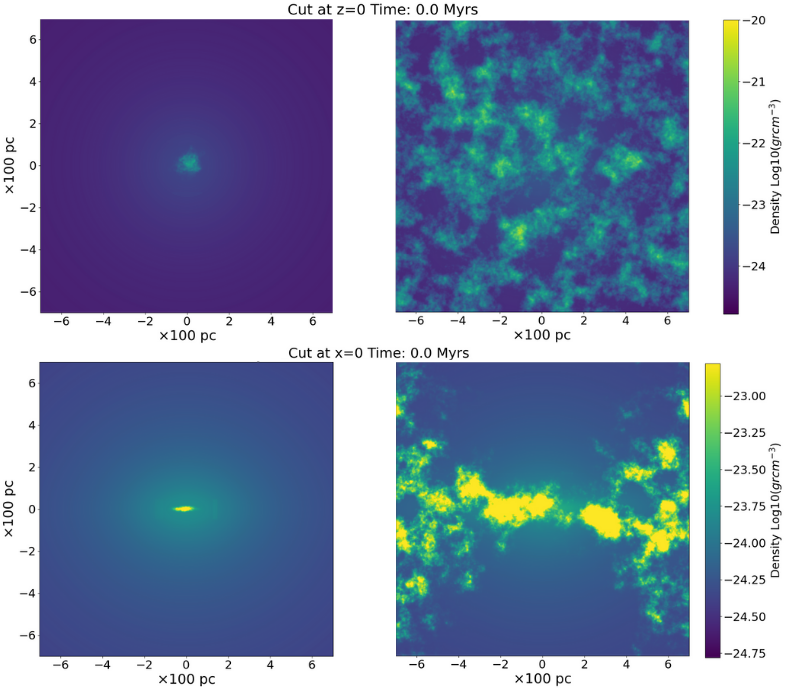}

  \caption{Density distribution from the top and side perspectives for two different models, one with $\sigma_t=0$ $\rm km\,s^{-1}$ (left column) and other with $\sigma_t=60$ $\rm km\,s^{-1}$ (right column). Top: simulation section at z=0. Bottom: simulation section at x=0. }
  \label{atmospheres}
\end{figure}

Figure \ref{atmospheres} depicts the density distribution from the top and side perspectives for two different models under adiabatic conditions, the first model has a value of the turbulence parameter $\sigma_t=0 \, \rm km\,s^{-1}$, while the second model has a value of $\sigma_t=60\, \rm km\,s^{-1}$.

On the other hand, our small-scale study involves the simulation of a two-phase ISM composed of spherical cold clouds surrounded by a hot supersonic wind with a uniform velocity distribution. This set-up can be viewed as a 3D section of the large-scale disc-wind models introduced above. The simulation domain is a rectangular prism with a 1:4:1 aspect ratio for the (W, L, H) domain where the width, length, and height of the prism are denoted as W, L, and H, respectively. To preserve all of the cloud material, periodic boundary conditions are included. 

Our small-scale wind-multicloud simulations cover a time-scale of $t_{\rm sim} \sim 5\,\rm Myr$, and include spherical clouds with identical radii $r_{\rm cl}=6.3\,\rm pc$, density $\rho_{\rm cl}=1.1 \times 10^{-24}\,\rm g\,cm^{-3}$, number density $n_{\rm cl}=1\rm cm^{-3}$, and temperature $T_{\rm cl}= 10^{4}\, \rm K$, placed vertically along the $y$ direction. The wind has a density of $\rho_{\rm wind}=1.1 \times 10^{-26}\,\rm g\,cm^{-3}$, number density $n_{\rm wind}=0.01 \rm cm^{-3}$, temperature $T_{\rm wind}=10^{6}\, \rm K$ and Mach number ${\cal M}_{\rm wind}=v_{\rm wind}/c_{\rm wind}=3.5$, where the wind speed is $v_{\rm wind}=500\,\rm km\,s^{-1}$ and its sound speed is $c_{\rm wind}=144\,\rm km\,s^{-1}$. The density contrast between cloud and wind gas is $\chi=10^2$, and both the wind and the multicloud system are initially in thermal pressure equilibrium at $P/k_{\rm b}= 10^{4}\,\rm K\,cm^{-3}$. The cloud centres are separated by different distances, $d_{\rm sep}$, which we report in units of the initial cloud radius, using a dimensionless parameter $\delta=d_{\rm sep}/r_{\rm cl}$. 

\subsection{Diagnostics}
Some diagnostics are computed from the data generated by our simulations, in order to asses the dynamical evolution of superbubbles and of a set of clouds in wind-multicloud systems. This analysis allows for a better understanding of the phenomena and the intrinsic mechanisms governing these systems. In the case of the formation of superbubbles, we define a cylindrical SF region with dimensions shown in Table \ref{table1}. Thus, the injection parameters, as mass and energy, can be calculated based on the injection equations of \cite{2008ApJ...674..157C}:

\begin{equation}
    \frac{dM}{dt \text{ }dV} = \frac{\dot{M}\rho}{\int \rho\text{ } dV}
    \label{mass_injection}
\end{equation}

\begin{equation}
    \frac{dE}{dt \text{ }dV} = \frac{\dot{E} \rho}{\int \rho\text{ } dV}
    \label{energy_injection}
\end{equation}

Where $\rho$ is the density integrated over the entire volume $V$, $\dot{M}$ and $\dot{E}$ are the mass and energy injection rate respectively. However, our simulations require that the energy be injected as thermal pressure (for code-related reasons). Therefore, with the help of the thermal equation of state we develop the following pressure injection equation:

\begin{equation}
    \frac{dP}{dt \text{ }dV} = \frac{\dot{E} \rho^2(\gamma-1)}{\left(\int \rho\text{ } dV\right)^2}
    \label{pressure_injection}
\end{equation}


On the other hand, to study the evolution of a set of clouds in a wind-multicloud model, several diagnostics can be calculated from the simulated data. we define the dense gas mass fraction \cite{heyer2022dense} as the ratio of cloud mass with a density greater than a threshold density $\rho'$ to the total cloud mass:
\begin{equation}\label{densefrac}
    f_{dense}(\rho>\rho') = \frac{M(\rho>\rho')}{M_{Tot}},
\end{equation}
where $\rho'$ is chosen based on the specific type of dense gas of interest. For this work, we are going to consider $\rho' = 0.5\rho_{cl,0}$ to focus on the cloud material that has more than half the density of the initial density value of the cloud. The mass of cold gas is determined by considering the mass of material with a temperature below $10\,T_{cl, 0}$, where $T_{cl, 0}$ represents the initial temperature of the cold gas. 
\begin{equation}\label{coldgassfrac}
    f_{cold} =  \frac{M(10T_{c}>T)}{M_{Tot}}.
\end{equation}
This corresponds to a temperature logarithmically halfway between the initial temperatures of the cold and hot media since the simulation is initialized in thermal pressure equilibrium with a density contrast of $\chi = 10^{2}$.

\section{Results}
\label{results}



\begin{figure}[htbp]
  \centering
  \includegraphics[width=1\linewidth]{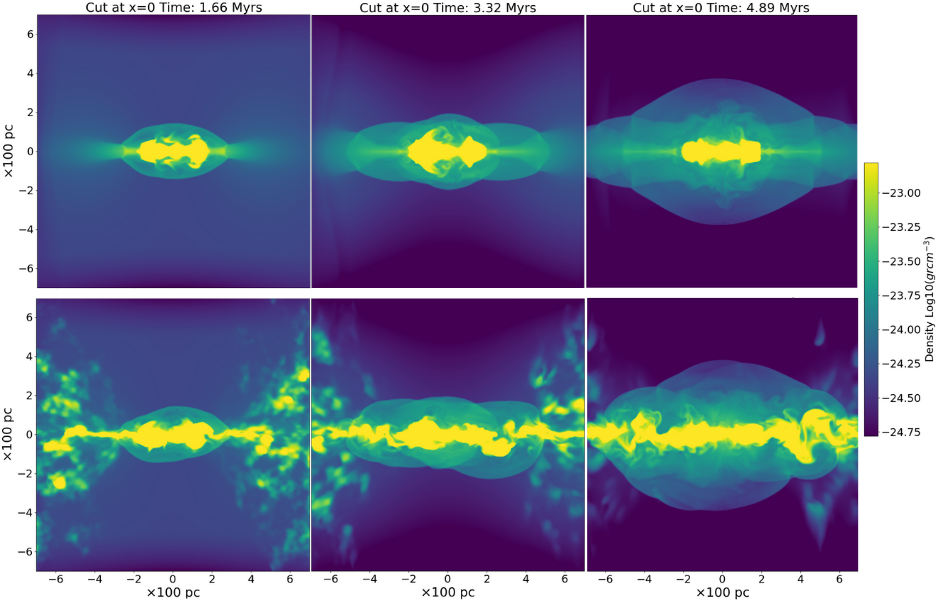} 

  \caption{Evolution of the superbubbles through the central plane ($x=0$) in a logarithm of density map. There are three times three models, the first without turbulence $\sigma_t=0$ $\rm km\,s^{-1}$ (top row), and the second with a thicker disc $\sigma_t=60$ $\rm km\,s^{-1}$ (bottom row).}
  \label{evolution_x_bubbles}
\end{figure}

In this section the evolution and formation of galactic winds in our simulations are described. We discuss the effect of changing some parameters in the models and how this change affects the morphology in the superbubble and in the wind-multicloud formation. Figure \ref{evolution_x_bubbles} shows the evolution of the large-scale simulations for three different times in a central plane.

Due to the release of energy and mass in the central region, it is possible to notice at 1.66 Myrs the outflows coming from the centre with densities on the order of $\sim 6\times10^{-24} \, \rm g\,cm^{-3}$. Because of the distribution of the gas in the first model with $\sigma_t=0$ $\rm km\,s^{-1}$, there is a presence of flows extended to the sides much more noticeable than in the other model. On the other hand, for $\sigma_t=60$ $\rm km\,s^{-1}$ the gas of the disc, which is distributed towards the sides prevents in some way the emergence of outflows along the plane. By 3.32 Myrs, the fall of the gas toward the GC by the action of gravity is more noticeable. Moreover, the gas pushed by the winds produced in the centre gradually reaches a higher altitude. For this time, there is a higher concentration of dense gas throughout the midplane for the second model than for the model without turbulence. Finally at 4.89 Myrs, the superbubbles have reached practically the same height in both models $(\sim 400 pc)$, however, for $\sigma_t=0$ model the larger bubble is more uniform and the dense gas is mostly concentrated in the injection region, while for the $\sigma_t=60$  $\rm km\,s^{-1}$ model, the gas is spread over the entire midplane and is thicker, reaching a height of about 100 pc. 

\begin{figure}[htbp]
  \centering
  \includegraphics[width=1\linewidth]{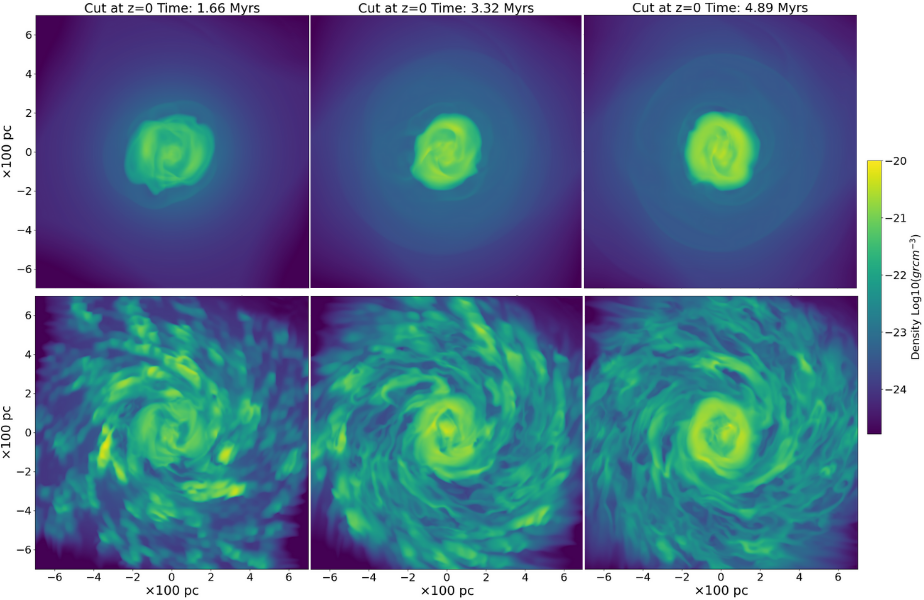} 

  \caption{The progression of superbubbles across the upper plane ($z=0$) is illustrated in a logarithmic density map. Three times are depicted from left to right, with the initial set having no turbulence ($\sigma_t=0$ $\rm km\,s^{-1}$) in the top row, and with turbulence ($\sigma_t=60$ $\rm km\,s^{-1}$) in the bottom row.}
  \label{evolution_z_bubbles}
\end{figure}

Figure \ref{evolution_z_bubbles} shows the evolution of the superbubble models seen from the upper plane. It is possible to observe the effect of disc rotation in both models. Nevertheless, the model with null turbulence parameter does not show a noticeable disc formation, while the other model for $\sigma_t=60$ $\rm km\,s^{-1}$ shows the presence of a well-formed disc. This is because the gas is distributed in the form of fractal clouds throughout the computational frame, not just in the center as in the first model.

Both models show the presence of outflows, which interact and merge giving rise to the formation of larger bubbles. In addition, this material, that is pushed out by the winds, reaches great heights ($\sim 400 $ pc) in agreement with observations in \cite{2019Natur.567..347P}. The low presence of dense gas to the north and south of the galactic plane facilitates the expansion of the gas in those directions. Nevertheless, for the model without turbulence the formation of a disc does not occur, unlike for the $\sigma_t=60,\rm km\,s^{-1}$ model. In the latter, the equilibrium between rotation and gravity is evident.

\begin{figure}[htbp]
  \centering
  \includegraphics[width=0.45\linewidth]{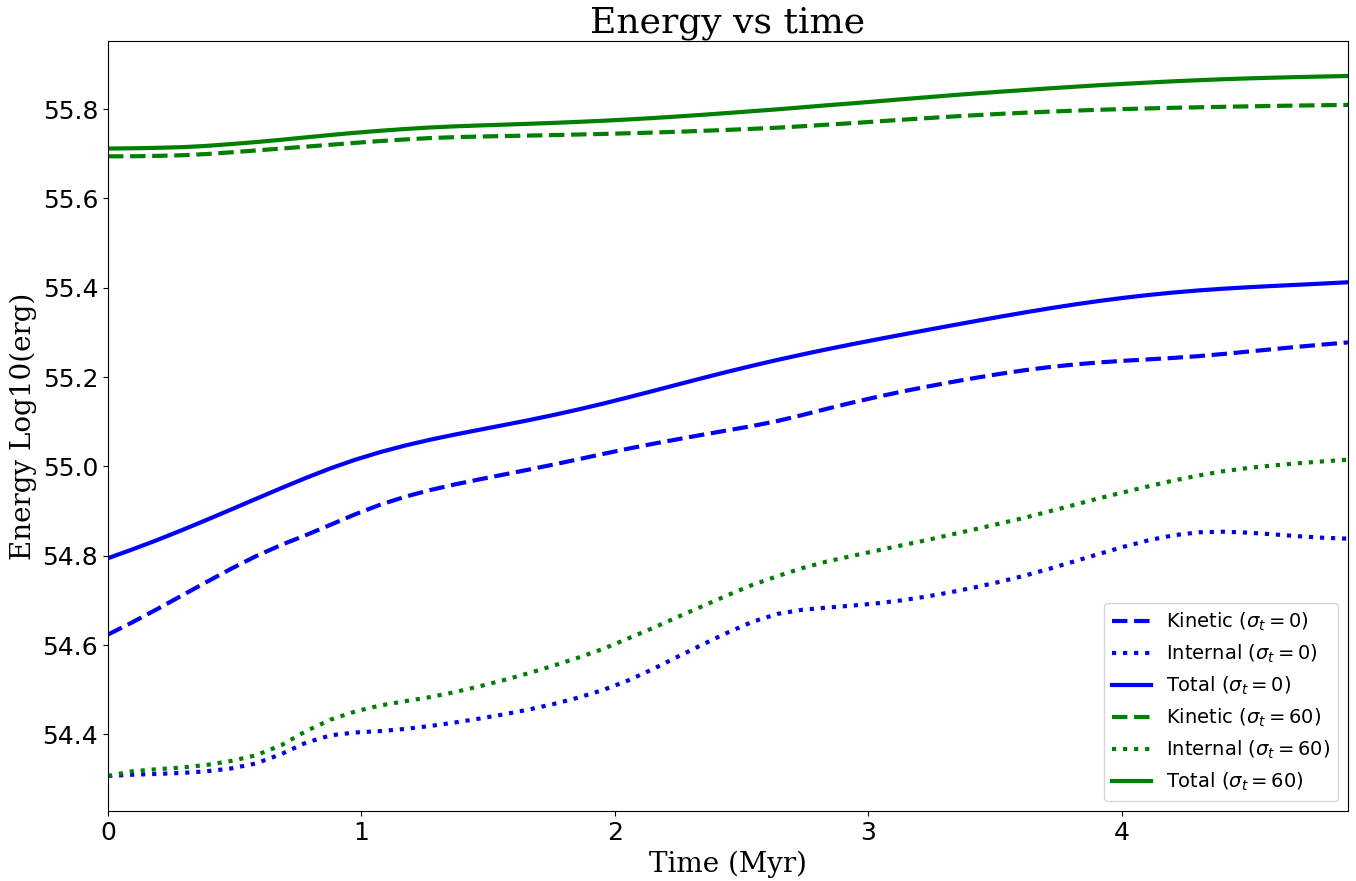} 
  \hspace{5pt}
  \includegraphics[width=0.45\linewidth]{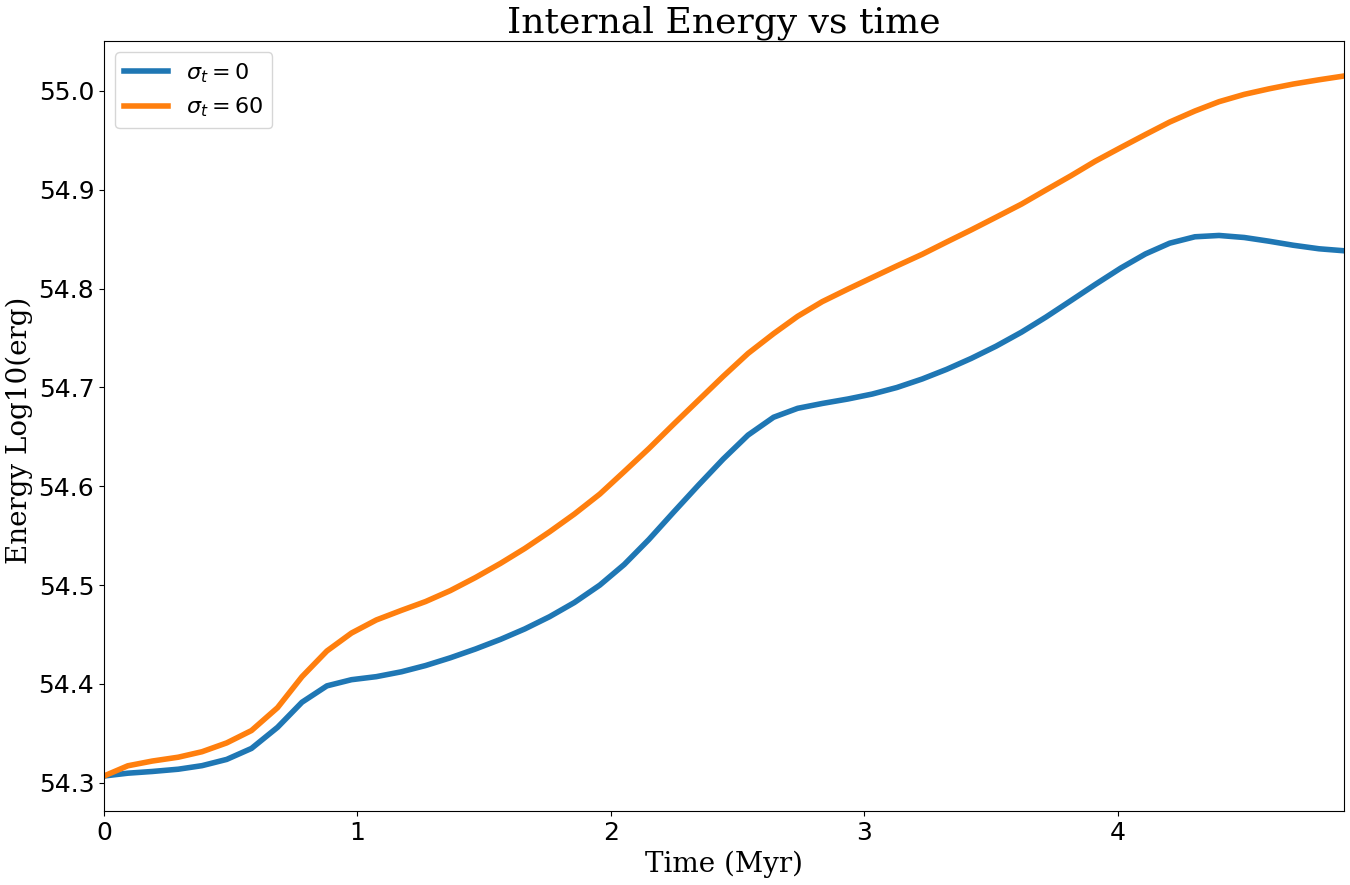}
  \vspace{5pt}
  \includegraphics[width=0.45\linewidth]{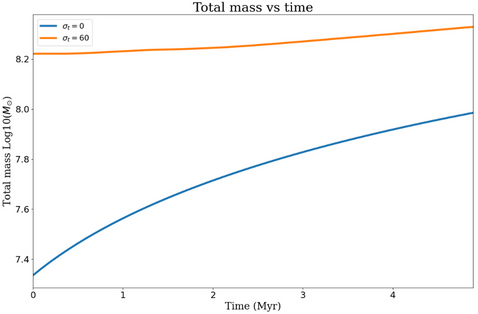}

  \caption{Time evolution of  the logarithm of total energy, kinetic energy and internal (top left). Logarithm of the internal energy changes over time (top right). Temporal evolution of the logarithm of total mass (bottom center).}
  \label{analysis_energy}
\end{figure}

Additionally, the turbulence parameter ($\sigma_t$) is important to control the scale height of the disc \cite{2008ApJ...674..157C}. However, this parameter is also important and can affect other physical quantities related to energetics. Figure \ref{analysis_energy} shows, in the top left side, the change in the logarithm of energy as time progresses. It is possible to notice that the major contribution for the total energy is the kinetic energy (around $5\times10^{55}$ erg for $\sigma_t=60$ $\rm km\,s^{-1}$ and $1.6\times10^{55}$ erg for $\sigma_t=0$ $\rm km\,s^{-1}$), which is related to what is mentioned in \cite{2000MNRAS.314..511S}. Followed to a lesser extent by the internal energy of the system with energies around $6\times10^{54}$ erg for $\sigma_t=0$ $\rm km\,s^{-1}$ and $8\times10^{54}$ erg for $\sigma_t=60$ $\rm km\,s^{-1}$. 

In the case of internal energy, see the image in the top right side of figure \ref{analysis_energy}, is possible to notice a periodicity in the increases and decreases of the energy in both models, this is due to the release of energies by the supernovae (crests)  occurring close to 1 Myr, 3 Myrs and slightly beyond 4 Myrs, and then a period of latency in which the gas accumulates giving origin to new stars (valleys) that later originate new explosions. Furthermore, the curve is growing as time passes, as it is related to the continuous injection of energy into the system as mentioned in \cite{2019Natur.567..347P}. 

In addition, in the model without turbulence, mass change occurs at a higher rate than in the $\sigma_t=60$ $\rm km\,s^{-1}$ model. The lower energy releases in the $\sigma_t=0$ $\rm km\,s^{-1}$ model result in more mass injections as time goes by, as less mass is expelled from the disc. In contrast, the other model, with more energetic outflows, drags more matter, resulting in a lower mass injection into the SF zone. Next we analyse how a small-scale 3D section of the global outflow evolves by studying wind-multicloud interactions.

\begin{figure}
\begin{center}
  \begin{tabular}{c c c c c c}
       \multicolumn{6}{l}{\hspace{-2mm}(a) Adiabatic ($\delta=2$)}\\
       \multicolumn{1}{c}{$t=0$} & \multicolumn{1}{c}{$t_{\rm sim}=0.4\,\rm Myr$} & \multicolumn{1}{c}{$t_{\rm sim}=1.1\,\rm Myr$} & \multicolumn{1}{c}{$t_{\rm sim}=1.9\,\rm Myr$} & \multicolumn{1}{c}{$t_{\rm sim}=2.6\,\rm Myr$} & \\   
       \hspace{-0.3cm}\resizebox{!}{67mm}{\includegraphics{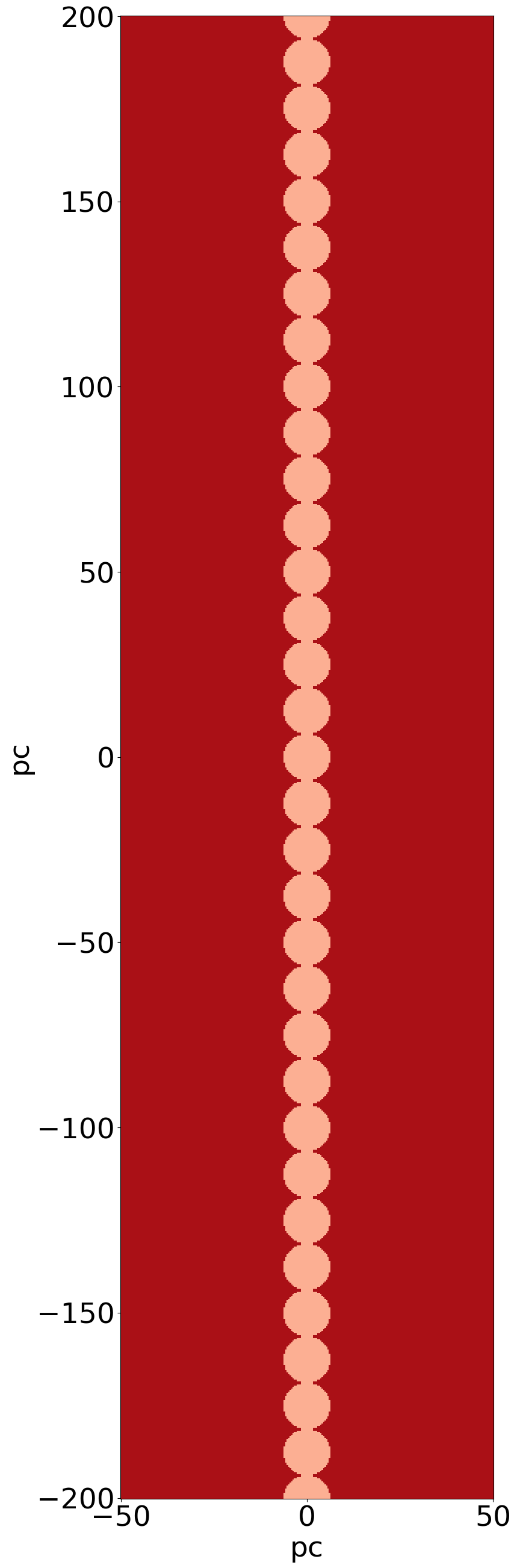}} & 
       \hspace{-0.3cm}\resizebox{!}{67mm}{\includegraphics{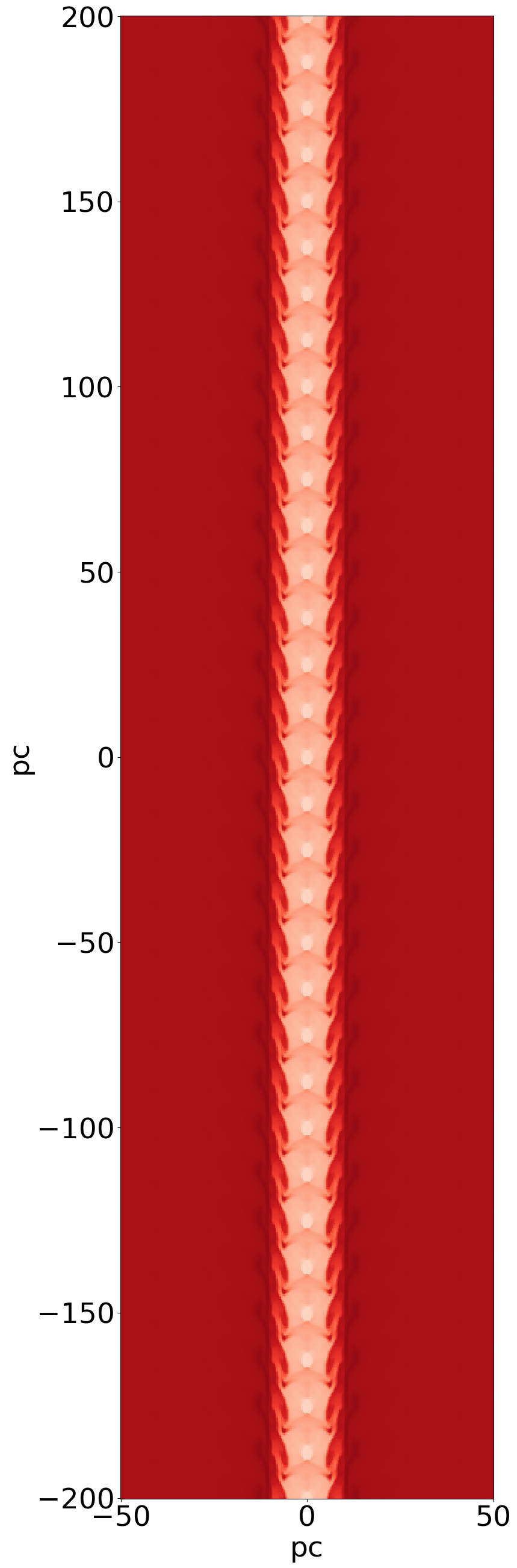}} & 
       \hspace{-0.3cm}\resizebox{!}{67mm}{\includegraphics{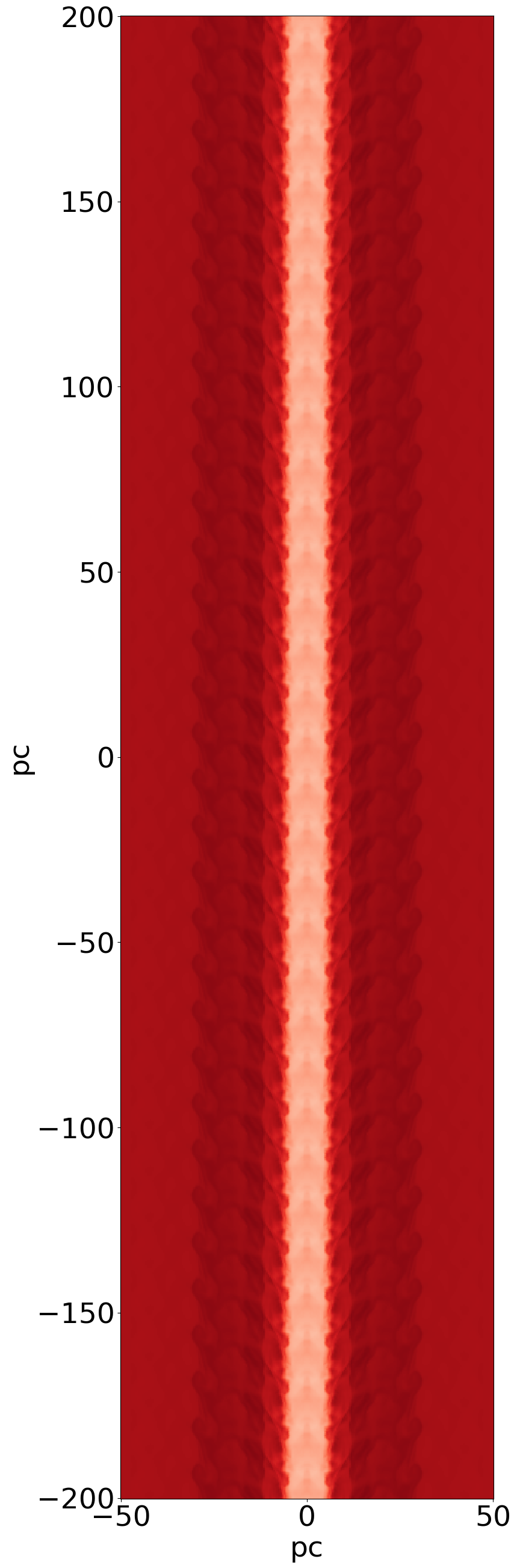}} & 
       \hspace{-0.3cm}\resizebox{!}{67mm}{\includegraphics{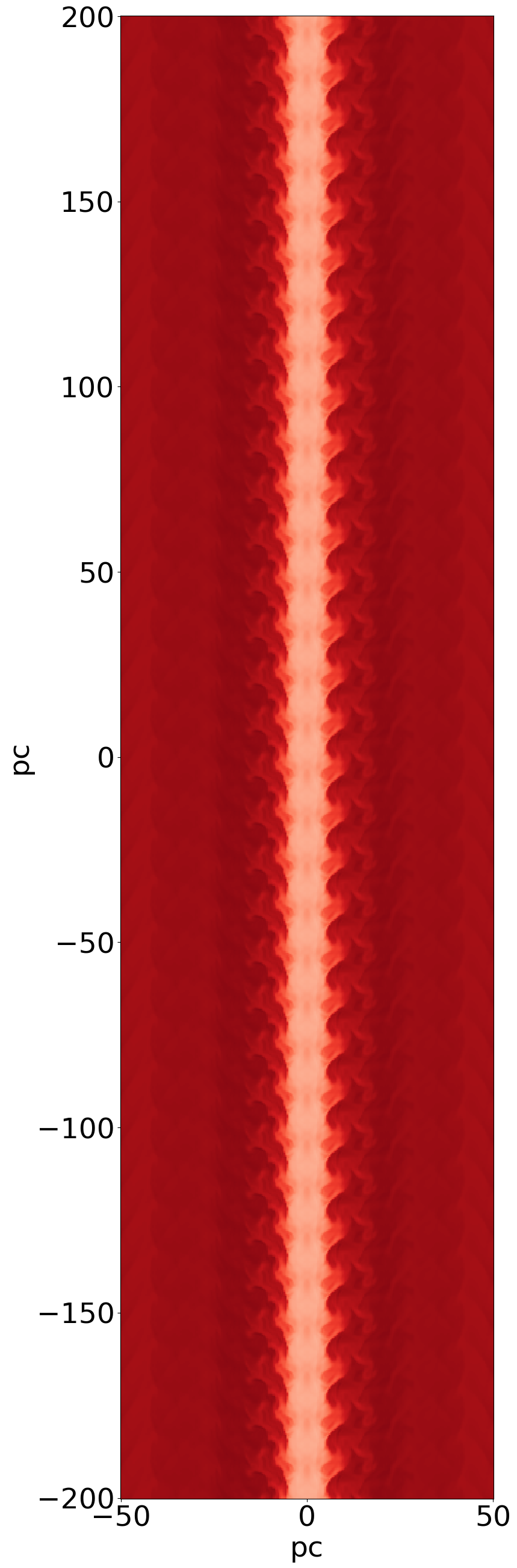}} & 
       \hspace{-0.3cm}\resizebox{!}{67mm}{\includegraphics{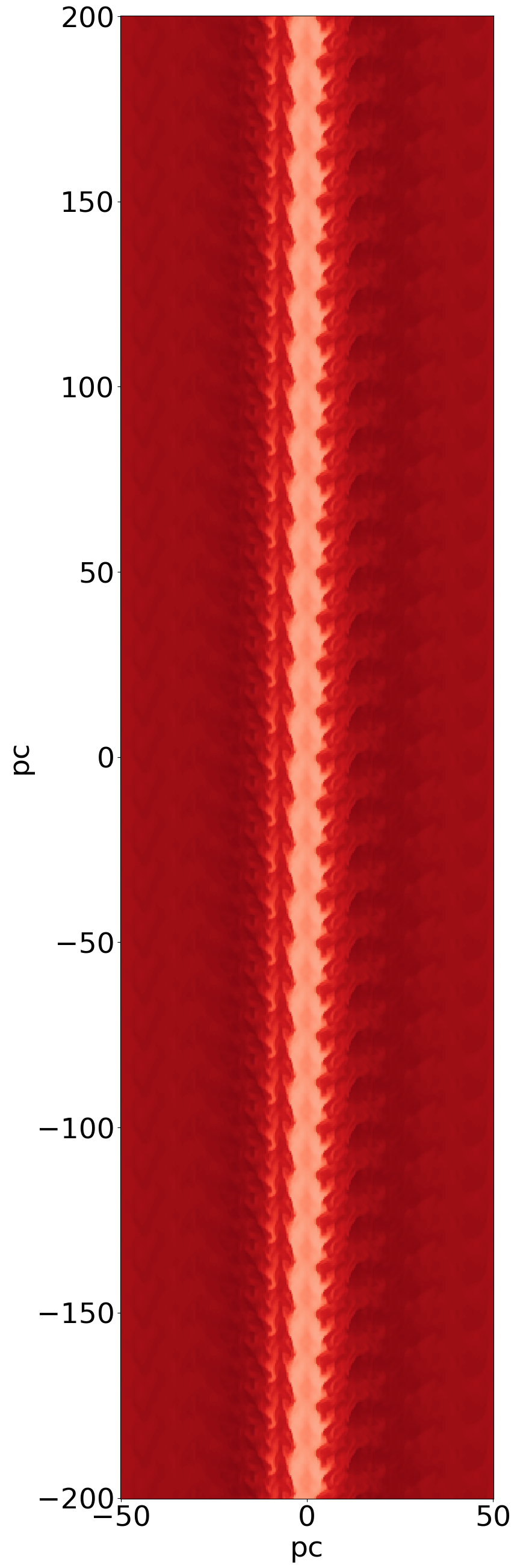}} \\
       \multicolumn{6}{l}{\hspace{-2mm}(b) Adiabatic ($\delta=16$)}\\
       \multicolumn{1}{c}{$t=0$} & \multicolumn{1}{c}{$t_{\rm sim}=0.4\,\rm Myr$} & \multicolumn{1}{c}{$t_{\rm sim}=1.1\,\rm Myr$} & \multicolumn{1}{c}{$t_{\rm sim}=1.9\,\rm Myr$} & \multicolumn{1}{c}{$t_{\rm sim}=2.6\,\rm Myr$} & \\     
       \hspace{-0.3cm}\resizebox{!}{67mm}{\includegraphics{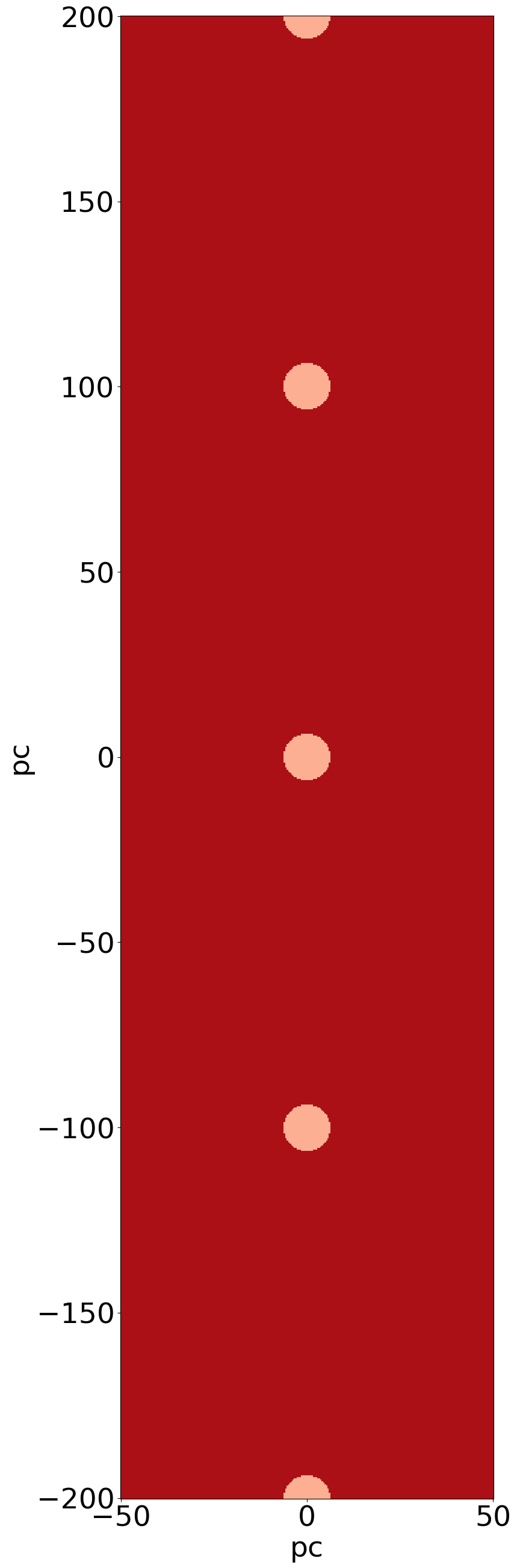}} & \hspace{-0.3cm}\resizebox{!}{67mm}{\includegraphics{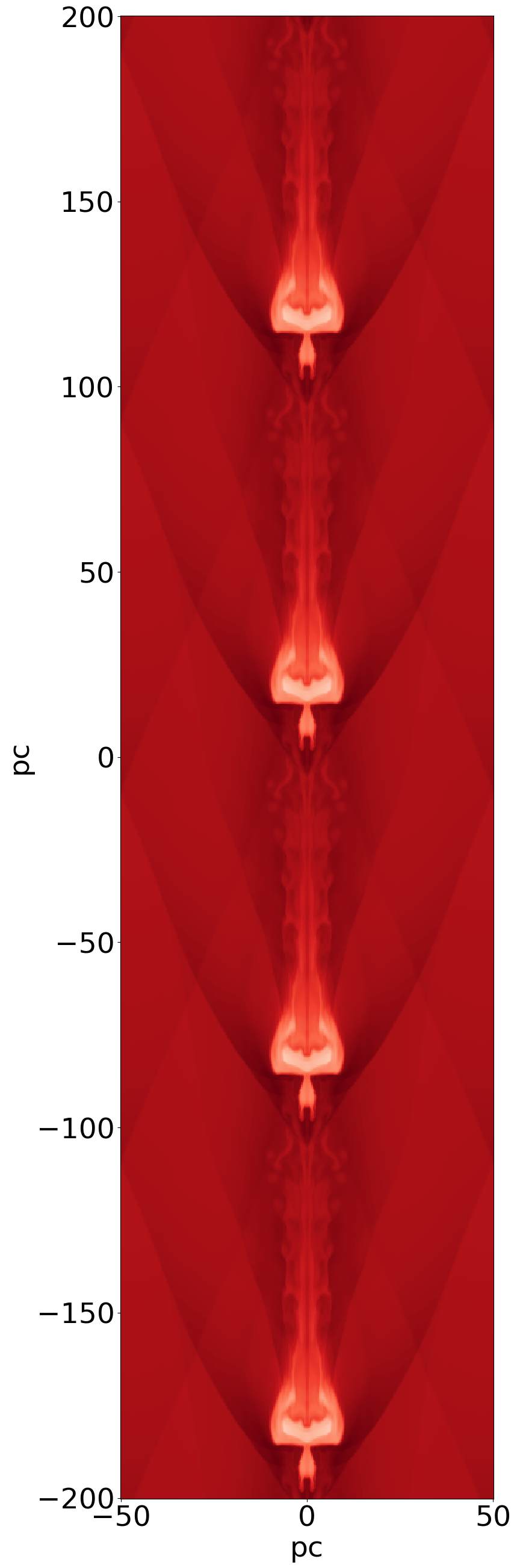}} & \hspace{-0.3cm}\resizebox{!}{67mm}{\includegraphics{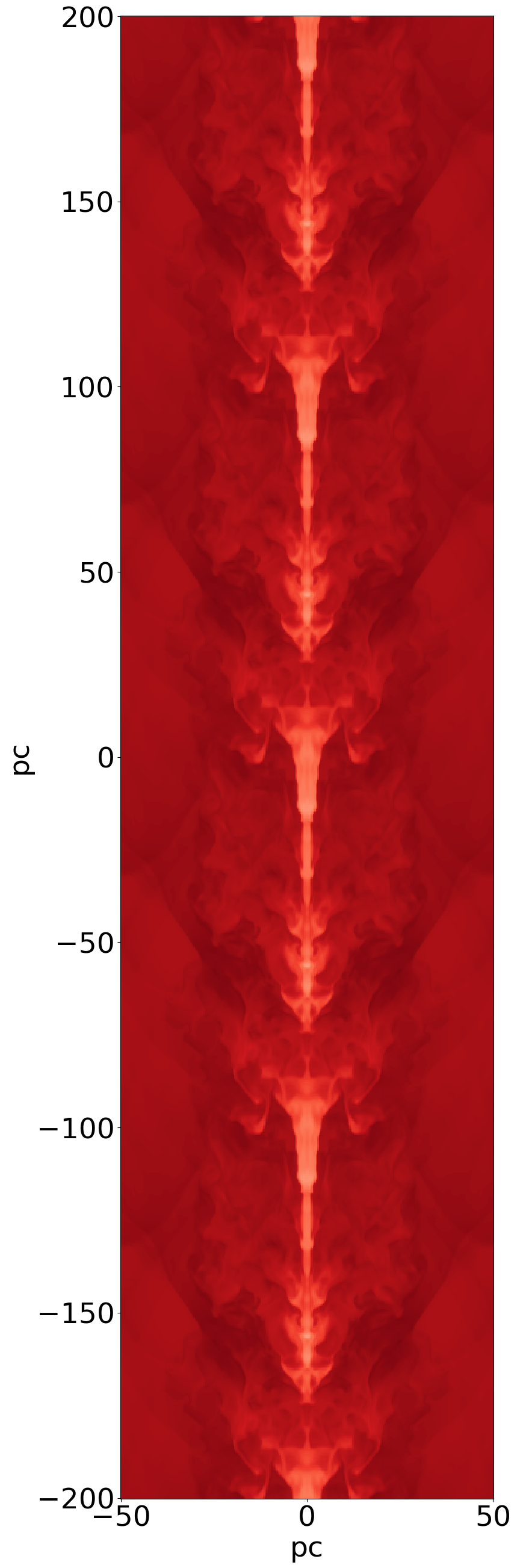}} & \hspace{-0.3cm}\resizebox{!}{67mm}{\includegraphics{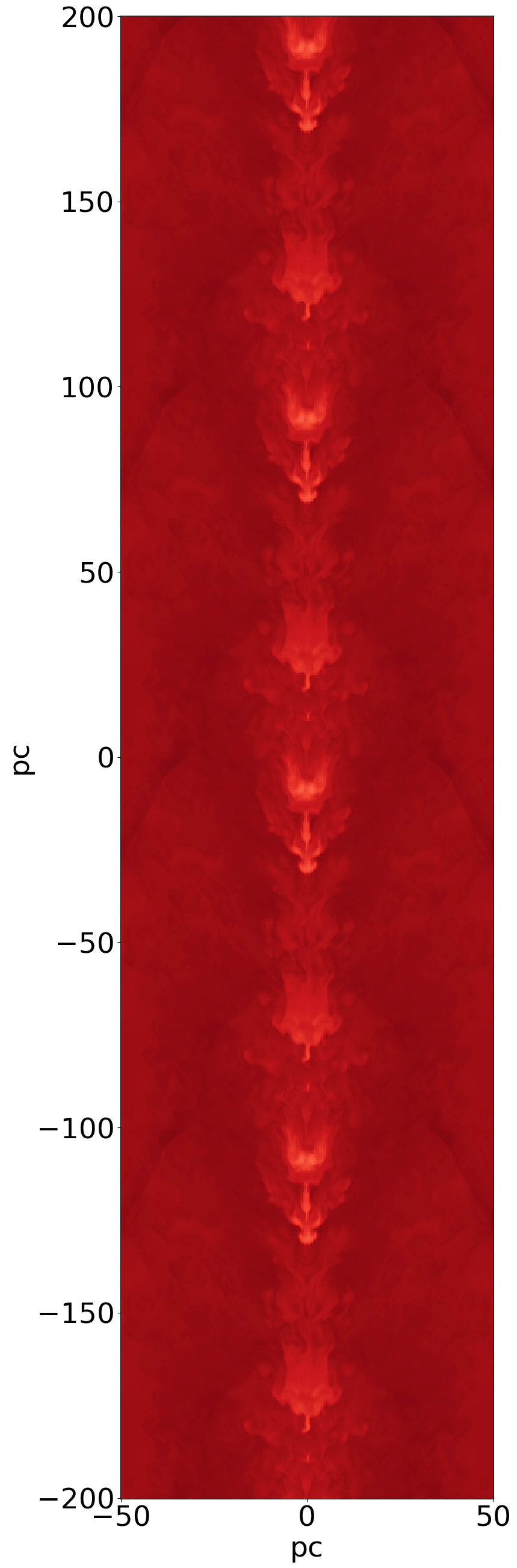}} & \hspace{-0.3cm}\resizebox{!}{67mm}{\includegraphics{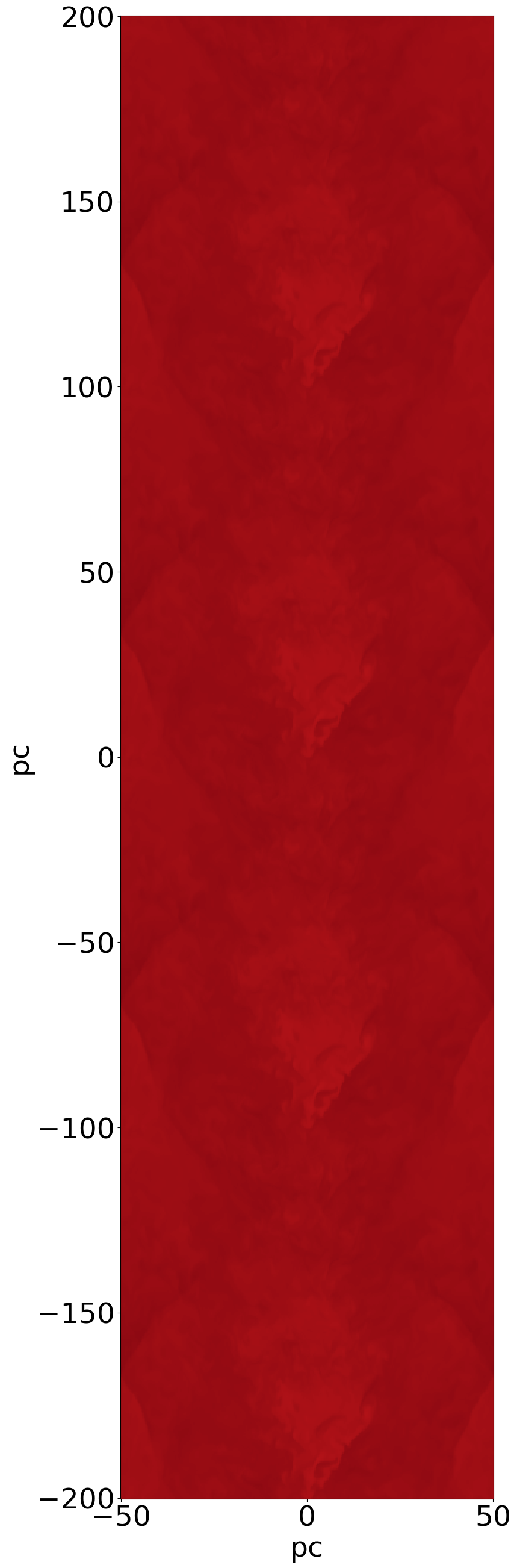}}\\
  \end{tabular}
  \begin{tabular}{c@{\hspace{-.035cm}}c}
  \end{tabular}
  \caption{2D slices at $Z=0$ of the gas temperature in adiabatic wind-multicloud models at five different times (shown in columns) throughout the simulation for different separation distances $\delta = 2$ (top panel a) and $\delta = 16$ (bottom panel b).}
  \label{temp1}
\end{center}
\end{figure}

Figure \ref{temp1} displays two-dimensional slices of the wind-multicloud models at Z = 0 of the gas temperature for different separation distances between clouds. The overall evolution of this pair of models can be characterized by the following stages: 1) The wind collides with the front of the clouds producing internal transmitted shocks accompanied by external reflected shocks; 2) Kelvin-Helmholtz (KH) instabilities emerge at the sides of the clouds and remove cloud gas; 3) The upstream motion of the wind triggers interactions between the gas removed from upstream clouds and that of downstream clouds; 4) The acceleration of clouds increases leading to the formation of Rayleigh-Taylor (RT) instabilities. These instabilities cause the destruction of clouds, as the cross-sectional area increases, which results in the disruption of the main cloud cores.

The vertical arrangement of clouds affects how they interact with the surrounding hot wind compared to cases where the clouds are isolated. We find that clouds in streams can shield themselves from disruption caused by the interaction with the supersonic wind gas. The interaction between clouds is a result of momentum transferred by the supersonic wind. 

Clouds separated by large distances ($\delta=16$), cannot withstand the hydrodynamic drag forces and completely disintegrate and mix with the wind. In contrast, multi-cloud setups with closely spaced clouds allow a large fraction of dense gas to remain shielded from the wind. Figure 6 shows that reducing the separation value between clouds results in a substantial increase in hydrodynamic shielding, which helps preserve the material of the cold cloud for a longer period of time.

\begin{figure*}
\begin{center}
  \begin{tabular}{r r} 
    \multicolumn{1}{l}{Dense gas mass fraction} & \multicolumn{1}{l}{Cold gas mass fraction}  \\ 
       \hspace{-0.3cm}\resizebox{!}{54mm}{\includegraphics{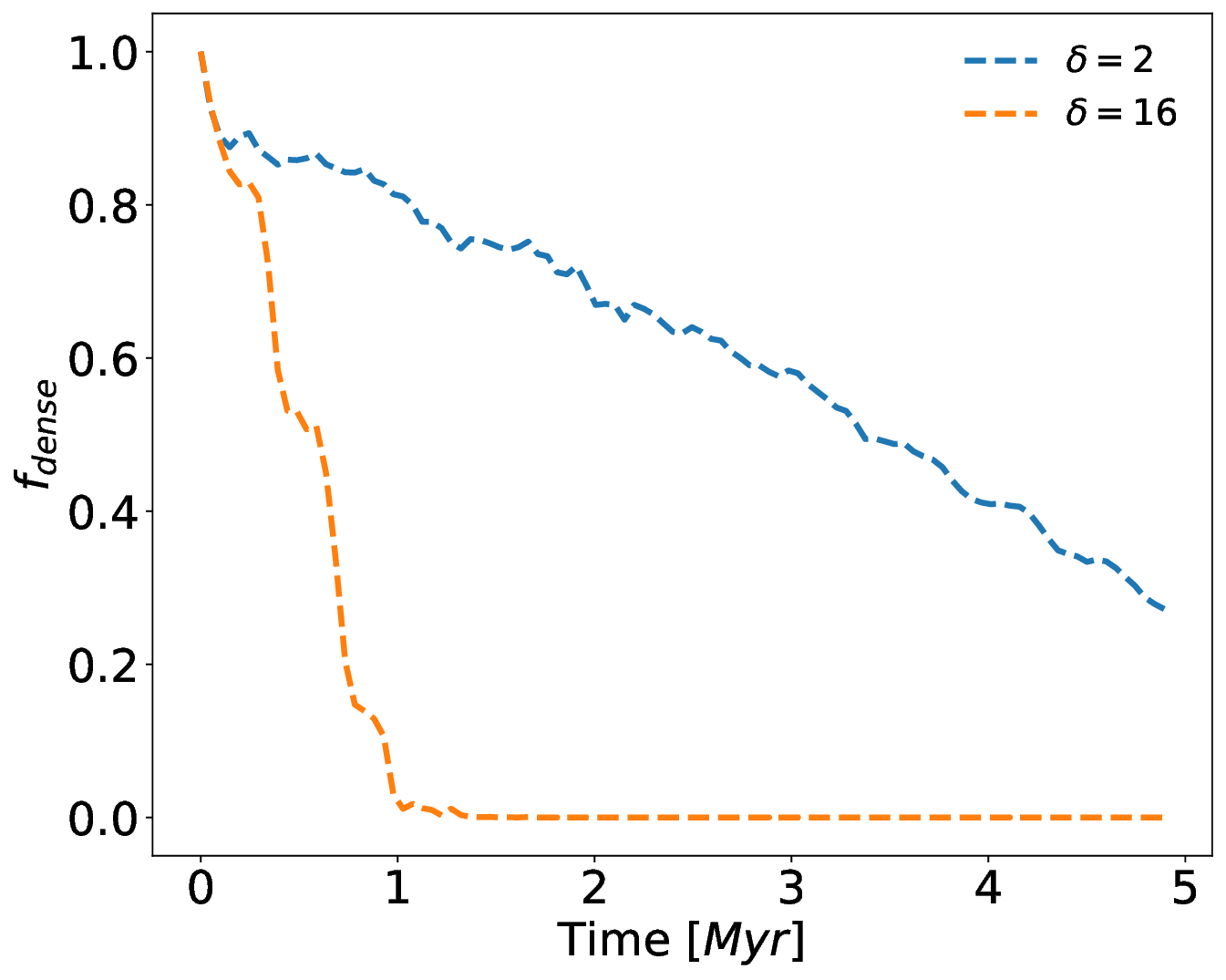}} & \hspace{-0.3cm}\resizebox{!}{54mm}{\includegraphics{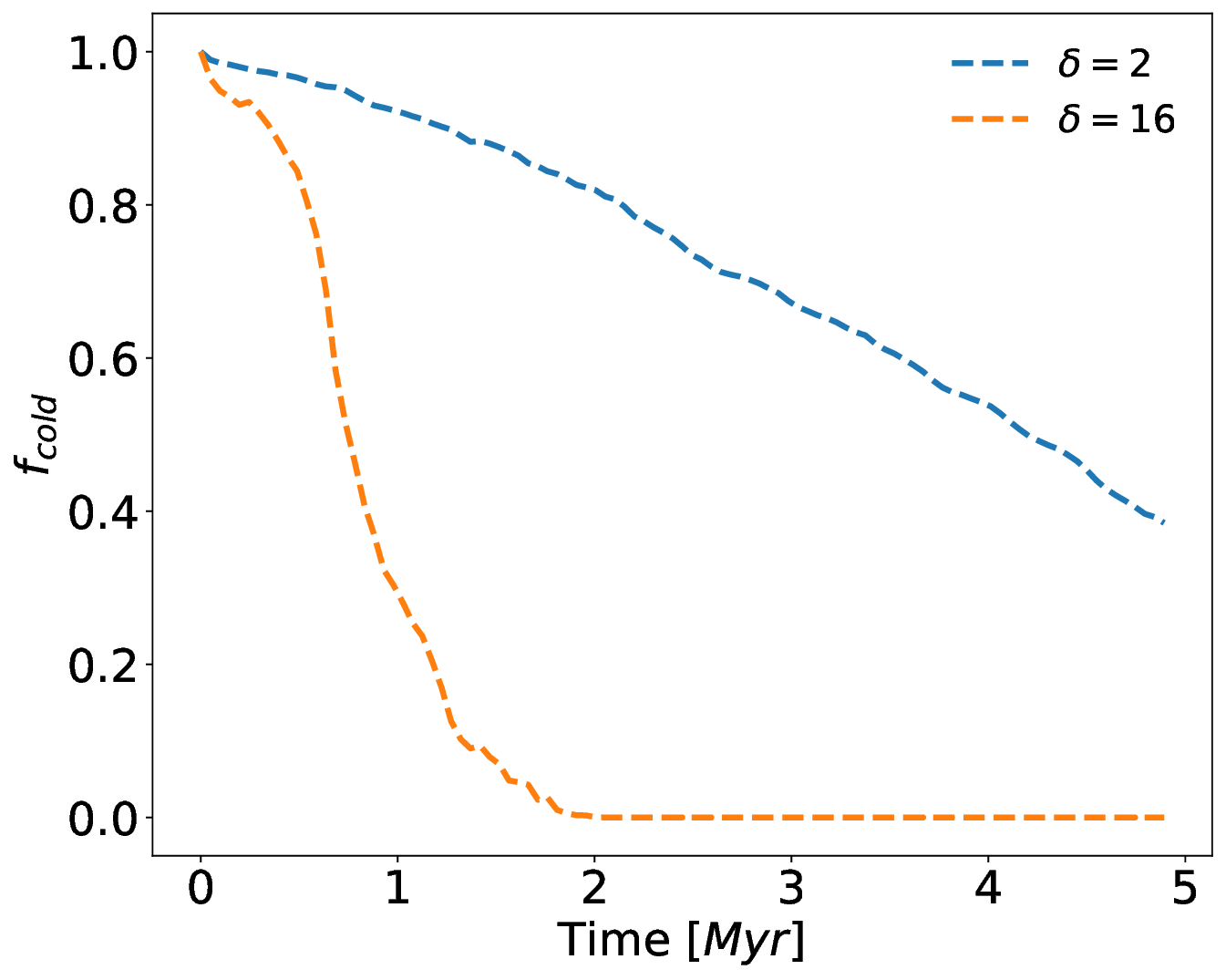}} 
  \end{tabular}
  \caption{Time evolution of the dense gas mass fraction (left panel) and the cold gas mass fractions (right panel) in the adiabatic model for different separation distances $\delta = 2, 16$} 
  \label{evo1}
\end{center}
\end{figure*}

Figure \ref{evo1} shows that the closely-spaced arrangement has a significant impact on adiabatic clouds (for the effects of hydrodynamic shielding on radiative clouds, please see \cite{Villares_etal2024}) as it offers protection to dense cloud cores by gas layers of intermediate-density material stripped from upstream clouds. Therefore, with the separation value $\delta = 2$ the arrangement conserves a dense gas mass of $>30 \%$ by 5 Myr. On the other hand, increasing the inter-cloud separation distance ($\delta$) results in a weaker hydrodynamic shielding effect, which leads to the quick destruction of the dense gas in the multicloud system by drag and instabilities.

Adiabatic clouds generally experience a rapid loss of their cold material at $t>0.9$ Myr, but hydrodynamic shielding is effective in small separation distances between clouds, enabling the clouds to survive twice as long as their counterparts. The hydrodynamic shielding mechanism proves to be particularly powerful in this regard, helping to maintain a fraction of cold gas throughout the entire simulation. 

\section{Conclusions}
\label{conclusions}

We have studied two types of simulations of the interface between the ISM and CGM, namely: disc-wind and wind-multicloud models. The first one is a large-scale model. In this case the GC is simulated so that itself-consistently gives rise to the formation of superbubbles. It was necessary to incorporate an inhomogeneous initial atmosphere in order to replicate the turbulent profile of the ISM disc\cite{2008ApJ...674..157C}. Furthermore, to facilitate the formation of superbubbles, it was necessary to introduce SN explosions. As these shells extend hundreds of parsecs above and bellow the Galactic plane, it was challenging to simulate each supernova. Consequently, a SF zone was modelled, where energy and mass were injected continuously as in the case of our Galaxy (e.g. \cite{2019Natur.567..347P}) and models of superbubbles (e.g. \cite{2008ApJ...674..157C}). Our main conclusions from these models are:

\begin{itemize}
    \item The key steps of the evolution of a superbubble as seen from a central plane are the following: first the gas is redistributed via gravity. Second, the injection zone replicates the supernova explosions in the GC resulting in outflows, which pushes certain amounts of gas. Third, the bubbles formed then merge reaching greater heights and giving rise to the formation of superbubbles. 

    \item The turbulence parameter influences disc formation, which is clearly noticed in the case of $\sigma_t=60$ $\rm km\,s^{-1}$ model, and reflects the equilibrium between the rotation velocity and the gravitational potential. On the other hand, for a $\sigma_t=0$ $\rm km\,s^{-1}$ model there are signs of gas rotation, but no noticeable disc is formed, only a high concentration of gas is observed in the injection zone.

    \item The turbulence parameter $\sigma_t$ not only controls the height scale of the disc, but also significantly influences the mass and energy characteristics of the system, leading to an overall increase in their magnitudes attributed to the control of $\sigma_t$ over the distribution of gas, thus, influencing the injection rate.

    \item The simulation recreates the conditions of a starburst-like zones. This is reflected in the periodicity of the internal energy curves, which periodically produces new winds that contribute to the formation of superbubbles.
\end{itemize}

Numerical simulations of wind-multicloud systems were also carried out as they can help understand the physical characteristics of the CGM gas across multiple scales. Through these simulations, we analysed the thermodynamic and turbulent properties of the different gas phases associated with the changing outflow (e.g. \cite{2021ApJ...913...68Z}, \cite{Forbes_2019}). We also conducted a series of 3D adiabatic hydrodynamical simulations of winds interacting with multi-cloud complexes. Here we presented only 2 cases, for further details the reader is referred to \cite{Villares_etal2024}. The simulations assessed the ability of systems of clouds, in vertical arrangements, to protect themselves against hydrodynamic drag and KH and RT instabilities arising from their interactions with a hot supersonic wind gas. Our conclusions from these models are:

\begin{itemize}
    \item Dynamic and thermodynamic differences arise when we change the initial separation distance between the clouds, $\delta$. The morphology of downstream clouds is not only influenced by the wind but also by the tails of upstream clouds that eventually collide with them. When the separation value is decreased, our models show that there is a significant increase in hydrodynamic shielding.
    \item Closely-spaced cloud arrangements have a substantial impact on the clouds, as they offer protection to dense and cold material from the instabilities. This material is preserved in stream-like structures, promoting the survival of dense gas.
\end{itemize}

\section*{Acknowledgments}
The authors gratefully acknowledge the Gauss Centre for Supercomputing e.V. (\url{www.gauss-centre.eu}) for funding this project by providing computing time (via grant pn34qu) on the GCS Supercomputer SuperMUC-NG at the Leibniz Supercomputing Centre (\url{www.lrz.de}). In addition, the authors thank CEDIA (\url{www.cedia.edu.ec}) for providing access to their HPC cluster as well as for their technical support. We also thank the developers of the PLUTO code for making this hydrodynamic code available to the community. Part of the work presented here is based on the thesis work of A. S. Villares (ASV) at Yachay Tech University (\cite{Villares_thesis}).

\end{document}